\documentclass[pra,aps,amsmath,amssymb,amsfonts,twocolumn,nofootinbib,longbibliography,superscriptaddress]{revtex4-2}
\usepackage{amsfonts}
\usepackage{txfonts}
\usepackage{amsmath}
\usepackage{float}
\usepackage{diagbox}
\usepackage[colorlinks,breaklinks,linkcolor=blue,anchorcolor=blue,citecolor=blue,urlcolor=blue]{hyperref}
\usepackage{graphicx}
\usepackage{dcolumn}
\usepackage{bm}
\usepackage{amssymb}
\usepackage{mathrsfs}
\usepackage[sort&compress]{natbib}
\usepackage{subfigure}
\usepackage{braket}
\usepackage{xcolor}

\usepackage{lineno}
 
\begin{document}
\title{Preparation of Multimode NOON States via Floquet-Engineered Conditional Chiral Excitation}

\author{Mengxue Li}
\affiliation{Center for Quantum Sciences and School of Physics, Northeast Normal University, Changchun 130024, China}

\author{Bo Tian}
\affiliation{Center for Quantum Sciences and School of Physics, Northeast Normal University, Changchun 130024, China}

\author{H. Z. Shen}
\affiliation{Center for Quantum Sciences and School of Physics, Northeast Normal University, Changchun 130024, China}

\author{Haodi Liu}
\affiliation{Center for Quantum Sciences and School of Physics, Northeast Normal University, Changchun 130024, China}

\author{Gangcheng Wang}
\email{wanggc887@nenu.edu.cn}
\affiliation{Center for Quantum Sciences and School of Physics, Northeast Normal University, Changchun 130024, China}

\date{\today}

\begin{abstract}
The NOON states possess Heisenberg-limited phase sensitivity and are crucial for quantum-enhanced metrology. In our work, we develop a Floquet-engineered conditional-routing
framework in which a $d$-level quantum controller mediates the
dynamics of $d$ bosonic modes. For an arbitrary odd dimension
$d=2s+1$, the $d$-level system can act as a ``quantum knob" by tuning its initial states to coherently control the direction of chiral excitation flow, such that different controller eigenstates generate distinct cyclic-routing branches of the bosonic excitations. By choosing $s$ harmonics, we construct an exact qudit-controlled cyclic permutation and use it to formulate a general protocol for preparing odd-$d$-mode NOON states. The same framework also enables programmable complex amplitudes through appropriate preparation of the controller state. As concrete realizations, we show that the three-mode case requires only a single harmonic and exhibits two oppositely directed chiral branches together with a stationary branch, whereas the five-mode case requires two harmonics and both nearest- and next-nearest-neighbor chiral hoppings. Meanwhile, taking the three-mode realization as a representative example, we assess the influence of systematic imperfections and dissipation on the conditional-routing dynamics and identify parameter regimes in which the routing operation remains accurate. Our results establish a systematic Floquet construction for controllable chiral routing and multimode NOON-state generation in arbitrary odd-dimensional networks, with potential applications in quantum information processing and quantum metrology.
\end{abstract}
\maketitle

\section{Introduction}
\label{Sec:I}
Entanglement, arising from the nonlocal superposition state of two or more quantum systems, is one of the most striking features of quantum mechanics \cite{RevModPhys.81.865,RevModPhys.90.035005,Gisin2007,Bennett2000}. As a key resource, quantum entanglement has attracted significant attention in quantum information processing \cite{NielsenChuang2010,S0097539795293172,rspa.1985.0070}, quantum communication \cite{RevModPhys.81.1301,Pirandola2015,qute.201900011,PhysRevA.65.032302,PhysRevLett.69.2881,PhysRevLett.81.5932,PhysRevLett.67.661}, and other related fields. Among various entangled states, NOON states \cite{science.1188172,ContempPhys.49.125,PhysRevLett.85.2733} represent a class of entangled states involving two orthogonal modes. The two-mode NOON state, denoted as $\ket{\rm{NOON}}=(\ket{N,0}+e^{i\varphi}\ket{0,N})/\sqrt{2}$, is commonly referred to as the photon number entangled state. This state represents a superposition between $N$ photons in one mode and zero photons in the other mode and vice versa. These states enable Heisenberg-limited scaling ($\sim 1/N$) in single-parameter phase estimation \cite{10.1116/5.0007577,PhysRevLett.129.020401,science.1138007,PhysRevLett.112.103604,PhysRevLett.97.150402,PhysRevA.93.043615} and have garnered significant attention in the field of quantum metrology \cite{RevModPhys.89.035002,RevModPhys.90.035005,PhysRevLett.122.040503,PhysRevLett.125.210503,PhysRevLett.132.190001,Panda2024}. 

Today, the concept of NOON states has been expanded to multimode NOON states, which are defined as the coherent superposition of $N$ photons in one mode and vacuum in all the other modes, to investigate quantum-enhanced precision in multiphase estimation \cite{PhysRevLett.111.070403,andp.202200304,Liu_2016}. Moreover, in the context of multiphase estimation under resource constraints \cite{Hong2021,lpor.202100682,Namkung_2024}, multimode NOON states have been demonstrated to enable enhanced sensitivity that outperforms both other quantum probe states and classical strategies in multimode interferometric setups \cite{PhysRevLett.111.070403,PhysRevA.106.032612,PhysRevA.95.032321,10.1116/5.0007577}, and are applicable to distributed sensing \cite{4vdx-7224}. Theoretical studies have also shown that three-mode NOON states enable the simultaneous estimation of two phase parameters with Heisenberg-limited precision \cite{Yao_2022}. Despite significant advancements, the high-fidelity preparation of NOON states still constitutes an unresolved challenge. Thus, sustained exploration of diverse approaches continues to hold considerable value for advancing the field.  More recently, multimode NOON-state preparation has been explored in interacting ultracold-atom systems using geodesic
counterdiabatic driving, which substantially accelerates the
adiabatic preparation process
\cite{qxqg-qnnq}.

In parallel, chiral flow represents a paradigmatic nonequilibrium dynamical phenomenon in quantum systems with time-reversal symmetry breaking, characterized by the directional and cyclic population transfer of excitations along closed paths or lattice structures \cite{PhysRevA.82.043811}. Breaking time-reversal symmetry through periodic modulation or synthetic gauge fields has enabled the implementation of such unidirectional transport in a variety of quantum systems, including Rydberg atoms \cite{PhysRevResearch.4.L032046,PhysRevA.109.032622}, trapped ions \cite{PhysRevA.97.010302}, giant atoms \cite{PhysRevA.107.023705}, cold atoms \cite{PhysRevA.110.012432}, photonic platforms \cite{PhysRevLett.126.103603} and circuit quantum electrodynamics \cite{PhysRevA.82.043811}. Relevant phenomena have also been experimentally verified in superconducting circuits \cite{Roushan2017,Wang2019,10.1063/1.5140884,Tao2021}. Owing to its unidirectional transport feature, chiral flow exhibits distinct advantages in quantum information processing, and can be utilized to develop nonreciprocal quantum routers \cite{PhysRevA.109.032622}, single-photon circulator \cite{PhysRevA.108.063715} and other key quantum devices. This unidirectional transport originates from the system’s global phase. For bosonic systems, modulated coupling phases can similarly enable chiral transport \cite{rspa.1984.0023,rsos.172447}. Such tunable chiral transport properties provide a feasible physical method for the preparation of multimode NOON states \cite{PhysRevApplied.23.054080}, and thus holds significant research value for constructing complex entangled states. 

In this work, we propose a scheme for preparing multimode NOON states based on conditional chiral excitation dynamics. Our approach is based on a periodically modulated multimode Jaynes–Cummings-type model. By deriving the effective Hamiltonian of the system, we realize conditional chiral excitation dynamics. We
introduce a multi-harmonic Floquet construction which synthesizes
long-range chiral hoppings and realizes an exact qudit-controlled
cyclic permutation for arbitrary odd dimension $d=2s+1$. Different initial states of the multilevel system can be chosen to induce distinct chiral dynamical evolutions. This crucial feature allows the multilevel system to act as a ``quantum knob” for tuning the network evolution. The $d=5$ case is presented explicitly as the first nontrivial realization beyond the original multimode architecture. It is directly applicable to a variety of platforms, including superconducting circuits and coupled photonic cavities \cite{RevModPhys.93.025005,PhysRevLett.113.083603}. 

The remainder of this paper is organized as follows.
In Sec.~\ref{Sec:II}, we develop the multi-harmonic Floquet construction and establish the corresponding qudit-controlled routing operation. In Sec.~\ref{Sec:III}, we formulate the general protocol for preparing odd-$d$-mode NOON states. In Sec.~\ref{Sec:IV}, we present the three- and five-mode realizations, illustrating respectively the single- and multi-harmonic constructions. In Sec.~\ref{Sec:V}, we investigate the sensitivity of the three-mode protocol against systematic imperfections and dissipation. In Sec.~\ref{Sec:VI}, we present two concrete experimental implementation schemes and discuss the feasibility of the proposed scheme. Finally, we summarize and conclude the entire work in Sec.~\ref{Sec:VII}.

\section{Floquet construction of controlled chiral dynamics}
\label{Sec:II}
\subsection{Multi-harmonic Floquet model}
\label{Sec:IIA}

\begin{figure}
\centering
\includegraphics[width=0.8\columnwidth]{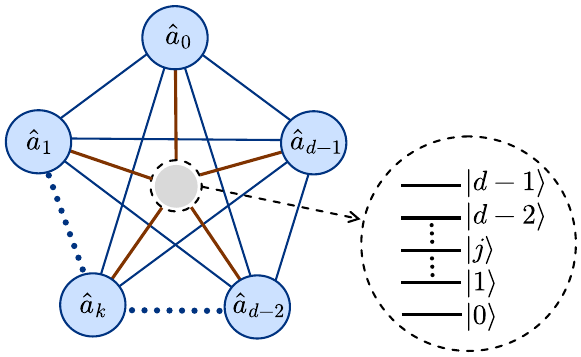}
\caption{Schematic diagram of the hybrid system consisting of a $d$-level system with adjacent level spacings $\omega_{q}$ and $d$ bosonic modes resonant at $\omega_{a}$, where the indirect coupling between bosonic modes is mediated by the $d$-level system.}
\label{Fig.1}
\end{figure}

We start our analysis with a composite system, consisting of an $d$-level subsystem and $d$ bosonic subsystems, as illustrated in Fig. \ref{Fig.1}. Throughout this work, we focus on an odd $d=2s+1$ with $s$ being a positive integer. The Hamiltonian of the composite system is ($\hbar\equiv1$)

\begin{equation}\label{eq.ori_H}
   \hat{H}(t) = \hat{H}_{0} + \hat{H}_{\rm int}(t),
\end{equation}
where
\begin{equation}
\begin{aligned}
   \hat{H}_{0} = &\sum_{j=0}^{d-1} \omega_{q_{j}}\ket{j}\bra{j}+\sum_{k=0}^{d-1}\omega_{a_{k}}\hat{a}_{k}^{\dagger}\hat{a}_{k},\\
   \hat{H}_{\rm int}(t)=&\sum\limits_{j=0}^{d-2}\sum_{k=0}^{d-1}g_{jk}(t)(\ket{j}\bra{j+1}\hat{a}_{k}^{\dagger}+\ket{j+1}\bra{j}\hat{a}_{k}).
\end{aligned}
\end{equation}
Here $\omega_{q_{j}}$ is the eigenenergy of the $j$th multilevel system, $\hat{a}_{k}$ $(\hat{a}_{k}^{\dagger})$ and $\omega_{a_{k}}$ are the annihilation (creation) operator and the frequency of the $k$th bosonic mode, respectively, and $g_{jk}(t)$ is the time-dependent coupling between the transition $\ket{j}\leftrightarrow\ket{j+1}$ and the $k$th bosonic mode. Here we assume only $d-1$ adjacent transitions are supported. Subsequent analyses are based on the assumption that the energy levels of the multilevel system are equally spaced, satisfying the relation $\omega_{q_{j}}=j\omega_{q}$ where $\omega_{q}$ denotes the energy level spacing, and that all bosonic modes have the same frequency and are resonant with the adjacent subsystem transitions $\omega_{a_{k}}=\omega_a=\omega_q$.

We now introduce the $s$ harmonics to engineer the desired conditional dynamics. Specifically, we periodically modulate the coupling as $g_{jk}(t)=\sum_{p=1}^{s}g_{j}^{(p)}\cos\left[p(\nu t+\varphi_{k})\right]$, where $p$ is the number of modulation harmonics and $\nu$ is the fundamental modulation frequency. The mode-dependent phases $\varphi_{k}=2\pi k/d$ denote the modulation phase. For the $p$th harmonic, we choose the transition-dependent coupling coefficient as $g_{j}^{(p)}=G_{p}\sqrt{j(d-j)}$, where $G_{p}$ determines the overall strength of the $p$th harmonic.

The particular $j$ dependence of $g_{j}^{(p)}$ allows the equally spaced $d$-level subsystem to be represented compactly by spin-$s$ operators. We therefore define the spin-$s$ operators as follows
\begin{equation}
\begin{aligned}
     \hat{S}_{+}&=\hat{S}_{-}^{\dagger}=\sum_{j=0}^{d-2}\sqrt{(j+1)(d-j-1)}\ket{j+1}\bra{j},\\
     \hat{S}_{z}&=\sum_{j=0}^{d-1}m_{j}\ket{j}\bra{j}.
\end{aligned}
\end{equation}
where $m_{j}=j-s$. These operators satisfy $[\hat{S}_{+},\hat{S}_{-}]=2\hat{S}_{z}$, $[\hat{S}_{z},\hat{S}_{\pm}]=\pm\hat{S}_{\pm}$. With these operators, the free Hamiltonian can be written as
\begin{equation}
\hat{H}_{0} = \omega_{q}\hat {S}_{z}+\sum_{k=0}^{d-1}\omega_{a_{k}}\hat{a}_{k}^{\dagger}\hat{a}_{k}. \label{eq:free_spin_representation} 
\end{equation}
An overall constant energy shift has been omitted from the Hamiltonian, as it contributes only a global phase to the time evolution. Under the resonance condition, the Hamiltonian in interaction picture is therefore
\begin{equation}
    \hat {H}_{I}(t)
    =
    \sum_{p=1}^{s}\sum_{k=0}^{d-1}
    G_{p}\cos\left[p(\nu t+\varphi_{k})\right]
    \left(
    \hat {S}_{+}\hat {a}_{k}
    +
    \hat {S}_{-}\hat {a}_{k}^\dagger
    \right).
    \label{eq:interaction_picture_spin}
\end{equation}
The spin-$s$ subsystem can be realized within the fully symmetric Dicke manifold of $2s=d-1$ identical two-level systems, with $\hat {\Sigma}_{\pm}=\sum_{\alpha=1}^{2s}\hat{\sigma}_{\alpha}^{\pm}$ and $\hat {\Sigma}_z=\frac{1}{2}\sum_{\alpha=1}^{2s}\hat{\sigma}_{\alpha}^{z}$. Provided that all two-level systems couple identically to each bosonic mode and are initially prepared in this symmetric manifold, the dynamics remains confined to its $d$-dimensional subspace. Our model is therefore equivalent to a $d$-mode Jaynes–Cummings model with multi-harmonic periodic modulation of the collective
light-matter couplings.

The interaction Hamiltonian can be expressed as a sum of Floquet components:
$\hat{H}_{I}(t) = \sum_{p=1}^{s} \left( \hat{H}_{p} e^{i p \nu t} + \hat{H}_{-p} e^{-i p \nu t} \right)$, where these components are given by
\begin{equation}
\hat H_{p}=\frac{G_{p}}{2}\sum_{k=0}^{d-1}e^{ip\varphi_{k}}
 \left(
    \hat {S}_{+}\hat {a}_{k}
    +
    \hat {S}_{-}\hat {a}_{k}^{\dagger}
    \right),
\end{equation}
with $\hat {H}_{-p}=\hat {H}_{p}^{\dagger}$.
In the high-frequency regime, the leading nonvanishing term of the van Vleck expansion is \cite{PhysRevX.4.031027,PhysRevLett.116.220502}:
\begin{equation}\label{eq.5}
\begin{aligned}
    \hat{H}_{\text{eff}} = &\sum_{p=1}^{s}\frac{[\hat {H}_{p},\hat {H}_{-p}]}{p\nu}\\
    =&\sum_{1\leq k<l\leq d}i\hat{S}_{z} J_{l-k}\left( \hat {a}_{k}^{\dagger} \hat {a}_{l} - \hat {a}_{l}^{\dagger} \hat {a}_{k} \right),
\end{aligned}
\end{equation}
where
\begin{equation}
    J_{l-k}=\sum_{p=1}^{s}\frac{G_{p}^{2}}{p\nu}
\sin\left[2\pi p(l-k)/d\right].
\end{equation}
Here we have combined the two oppositely ordered mode pairs $(k,l)$ and $(l,k)$. For the odd dimension $d=2s+1$, we then introduce the cyclic separation $r=l-k$ (mod $d$). The effective Hamiltonian can be rewritten as follows.
Consequently, we obtain
\begin{equation}
 \hat {H}_{\mathrm{eff}}
 =\hat {S}_{z}\otimes\hat {H}_{\mathrm{cyc}},
 \label{eq:Heff_Sz_Hcyc}
\end{equation}
where the purely bosonic cyclic generator is
\begin{equation}
 \hat {H}_{\mathrm{cyc}}
 =i\sum_{k=0}^{d-1}\sum_{r=1}^{s}
 J_{r}
 \left(
 \hat {a}_{k}^{\dagger}\hat a_{[k+r]_{d}}
 -\hat {a}_{[k+r]_{d}}^{\dagger}\hat {a}_k
 \right),
 \label{eq:Hcyc_realspace}
\end{equation}
with $[n]_d$ denotes $n$ modulo $d$ with values in $\{0,\dots,d-1\}$. Here $r=1,\dots,s$, so that every pair of distinct bosonic modes is counted once. 
The above effective Hamiltonian is found to exhibit intriguing conditional dynamics, in which the evolution of the bosonic modes is dependent on the quantum state of the multilevel system. 

The controller eigenvalue $m_j$ determines not only the direction
but also the characteristic propagation velocity of the conditional
chiral dynamics. To see this explicitly, we introduce the Fourier
modes
\begin{equation}
\hat{\tilde a}_{q_n}
=
\frac{1}{\sqrt d}
\sum_{k=0}^{d-1}
e^{-iq_n k}\hat a_k,
\end{equation}
in terms of which the cyclic Hamiltonian has the dispersion $\varepsilon(q_n)=-2\sum_{r=1}^{s}J_r\sin(rq_n)$ with $q_n=2\pi n/d$.
For the controller state $\ket{j}$, the corresponding dispersion is
$\varepsilon_j(q_n)=m_j\varepsilon(q_n)$. To characterize the propagation velocity, we regard the discrete
spectrum as the sampling of the continuous dispersion $\varepsilon_j(q)=-2m_j\sum_{r=1}^{s}J_r\sin(rq)$ at the allowed momenta $q=q_n$, and hence the group velocity is
\begin{equation}
v_g^{(j)}(q)
=
\frac{\partial\varepsilon_j(q)}{\partial q}
=
-2m_j
\sum_{r=1}^{s}
rJ_r\cos(rq).
\label{eq:group_velocity}
\end{equation}
Thus, the sign of $m_j$ reverses the propagation direction, whereas
its magnitude rescales the propagation velocity. In particular, the
$m_j=\pm2$ sectors propagate twice as fast as the $m_j=\pm1$ sectors.
This is also evident from
$U_j(t)=\exp(-im_jH_{\rm cyc}t)=U_{1}(m_jt)$.

Equation~\eqref{eq.5} satisfies several symmetry relations: First, it obeys the excitation number conservation with $[\hat{H}_{\rm{eff}},\hat{N}]=0$, where $\hat{N}=\sum_{k=0}^{d-1}\hat{a}_{k}^{\dagger}\hat{a}_{k}$ is the bosonic excitation number operator, which ensures that for negligible dissipation, excitations are neither spontaneously generated nor annihilated during system evolution, with only coherent transfer of excitations between modes occurring. This excitation conservation is a direct manifestation of the global $U(1)$ symmetry inherent to the system. In addition, consider the time-reversal operator $\mathcal{T}=\mathcal{K}$, where $\mathcal{K}$ denotes complex conjugation. Due to the anti-unitary nature of 
$\mathcal{T}$, it conjugates complex numbers $\mathcal{T}i\mathcal{T}^{-1}=-i$. Therefore, the full Hamiltonian transforms under time reversal as $\mathcal{T}\hat{H}_{\rm{eff}}\mathcal{T}^{-1}=-\hat{H}_{\rm{eff}}\neq \hat{H}_{\rm{eff}}$, breaking the time-reversal symmetry, which is the essential physical origin of the chiral cyclic evolution of the bosonic modes observed in our scheme.

\subsection{Qudit-controlled cyclic permutation}
\label{Sec:IIB}
We now examine the resulting dynamics and show that an appropriate
choice of the harmonic amplitudes realizes the desired cyclic shift
at a common evolution time. To express the effective Hamiltonian
in matrix form, we introduce the cyclic shift matrix
\begin{equation}
    \mathcal{M}_{d} =
    \begin{bmatrix}
        0 & 1 & 0 & \cdots & 0 \\
        0 & 0 & 1 & \cdots & 0 \\
        \vdots & \vdots & \vdots & \ddots & \vdots \\
        0 & 0 & 0 & \cdots & 1 \\
        1 & 0 & 0 & \cdots & 0
    \end{bmatrix},
    \label{eq:cyclic_shift_matrix}
\end{equation}
which satisfies
$\mathcal{M}_{d}^{d}=I_{d}$ and
$\mathcal{M}_{d}^{\dagger}=\mathcal{M}_{d}^{-1}$.
Acting $\mathcal{M}_{d}$ on the mode-operator vector
$\hat{\bm A}=(\hat {a}_{0},\hat {a}_{1},\ldots,\hat {a}_{d-1})^{\mathsf {T}}$ with $\mathsf {T}$ denote matrix transposition, we obtain $\mathcal{M}_{d}\hat{\bm {A}}=(\hat {a}_{1},\hat {a}_{2},\ldots,\hat {a}_{d-1},\hat {a}_{0})^{\mathsf {T}}$.
More generally, $(\mathcal{M}_{d}^{r}\hat{\bm {A}})_k=\hat {a}_{[k+r]_{d}}$,
with mode indices understood modulo $d$. It follows that $\hat{\bm A}^{\dagger} \mathcal{M}_{d}^{-r}\hat{\bm {A}}=\sum_{k=0}^{d-1}\hat {a}_{[k+r]_{d}}^{\dagger}\hat {a}_{k}$, and $\hat{\bm {A}}^{\dagger} \mathcal{M}_{d}^{r}\hat{\bm {A}}=\sum_{k=0}^{d-1}\hat {a}_{k}^{\dagger}\hat {a}_{[k+r]_{d}}$.
Using these identities, Eq.~\eqref{eq:Hcyc_realspace}
takes the quadratic form $\hat {H}_{\mathrm{cyc}}=\hat{\bm {A}}^{\dagger} h_{\mathrm{cyc}}\hat{\bm {A}}$, where the matrix $h_{cyc}$ reads
\begin{equation}
    h_{\mathrm{cyc}}
    =
    i\sum_{r=1}^{s}
    J_{r}\left(\mathcal{M}_{d}^{r}-\mathcal{M}_{d}^{-r}\right).
    \label{eq_hcyc}
\end{equation}
Consequently, Eq.~\eqref{eq:Heff_Sz_Hcyc} becomes
\begin{equation}
    \hat {H}_{\mathrm{eff}}
    =
    \hat {S}_{z}\hat{\bm {A}}^{\dagger}
    h_{\mathrm{cyc}}\hat{\bm {A}}.
\end{equation}
The Heisenberg equation of motion for the mode-operator vector
$\hat{\bm {A}}(t)$ is
\begin{equation}
    \frac{d\hat{\bm {A}}(t)}{dt}
    =
    -i\hat {S}_{z} h_{\mathrm{cyc}}\hat{\bm {A}}(t).
    \label{eq:Heisenberg_equation}
\end{equation}
Since $[\hat {S}_{z},\hat {H}_{\mathrm{eff}}]=0$ and the effective
Hamiltonian is time independent, the solution reads
\begin{equation}
    \hat{\bm {A}}(t)
    =
    \hat {U}^{\dagger}(t)\hat{\bm {A}}(0)\hat {U}(t)
    =
    \hat {R}(t)\hat{\bm {A}}(0),
    \label{eq:Heisenberg_solution}
\end{equation}
where $ \hat {U}(t)=e^{-i\hat {H}_{\mathrm{eff}}t}$ and $\hat {R}(t)=e^{-i(\hat {S}_{z}\otimes h_{\mathrm{cyc}})t}$. Using the controller basis $\{|j\rangle,j=0,\dots,d-1\}$,
with $\hat {S}_{z}| j\rangle=m_{j}| j\rangle$ and
$m_{j}=j-s$, we obtain the block decomposition
\begin{equation}
    \hat {R}(t)
    =
    \sum_{j=0}^{d-1}
    | j\rangle\langle j|\otimes R_{j}(t),
    \label{eq:conditional_evolution}
\end{equation}
where $R_{j}(t)=e^{-im_{j}h_{\mathrm{cyc}}t}$ describes the conditional mode evolution when the controller is in state $| j\rangle$.

To reveal the routing dynamics generated by the effective Hamiltonian derived above, we introduce a positive frequency scale $\Omega$ that
sets the quasienergy spacing of the engineered cyclic generator and the characteristic routing time, and choose $G_{p}=p\sqrt{2\Omega\nu/d}$. For odd $d$, the finite sine sum can be evaluated exactly
\begin{equation}
J_{r}=\frac{\Omega}{2}(-1)^{r+1}
\csc\left(\frac{\pi r}{d}\right).
\label{eq:Jr_closed}
\end{equation}
Defining $\theta_{j}(t)=m_{j}\Omega t$ and expanding the conditional evolution matrix in powers of
$\mathcal {M}_{d}$, we obtain
\begin{equation}
    R_{j}(t)
    =
    \frac{1}{d}
    \sum_{r=0}^{d-1}
    \alpha_{r}^{(j)}(t)\mathcal {M}_{d}^{r},
    \label{eq:Rj_expansion}
\end{equation}
where
\begin{equation}
    \alpha_{r}^{(j)}(t)
    =
    1+2\sum_{n=1}^{s}
    \cos\left[
        n\left(\theta_{j}(t)-\frac{2\pi r}{d}\right)
    \right].
    \label{eq:alpha_general}
\end{equation}
Inserting the matrix elements $(\mathcal M_d^r)_{kl}=\delta_{k,l-r}$ (indices modulo $d$) into Eq.~\eqref{eq:Rj_expansion}, we arrive at the closed form
\begin{equation}
    [R_j(t)]_{kl}=\frac{1}{d}\left[1+2\sum_{n=1}^s\cos \left(n\theta_j(t)+\frac{2\pi n(k-l)}{d}\right)\right].
\end{equation}
At the common routing time $\tau_d=2\pi/(d\Omega)$,
these coefficients satisfy
$\alpha_r^{(j)}(\tau_d)=d\delta_{r,[m_j]_d}$,
yielding $R_j(\tau_d)=\mathcal M_d^{m_j}$.

We next establish the corresponding evolution operator on the
joint controller--bosonic Hilbert space. Since
$\hat H_{\mathrm{eff}}=\hat S_z\otimes\hat H_{\mathrm{cyc}}$,
the spectral decomposition of $\hat S_z$ gives
\begin{equation}
    \hat H_{\mathrm{eff}}
    =
    \sum_{j=0}^{d-1}
    | j\rangle\langle j|
    \otimes m_j\hat H_{\mathrm{cyc}}.
\end{equation}
The mutually orthogonal controller projectors therefore yield
\begin{equation}
    \hat U(t)
    =
    e^{-i\hat H_{\mathrm{eff}}t}
    =
    \sum_{j=0}^{d-1}
    | j\rangle\langle j|
    \otimes\hat U_j(t),
    \label{eq:Ueff_general}
\end{equation}
where $\hat U_j(t) = e^{-im_j\hat H_{\mathrm{cyc}}t}$. Here $\hat U_j(t)$ acts on the bosonic Fock space. According to Eq.~\eqref{eq:Heisenberg_solution}, we obtain
\begin{equation}
    \hat U_j^\dagger(t)\hat a_k\hat U_j(t)
    =
    \sum_{l=0}^{d-1}[R_j(t)]_{kl}\hat a_l=\frac{1}{d}\sum_{r=0}^{d-1}\alpha_{r}^{(j)}(t)\hat{a}_{[k+r]_{d}}.
    \label{eq:Uj_Rj_relation}    
\end{equation}
Equivalently,
\begin{equation}
     \hat U_j(t)\hat a_k\hat U^\dagger_j(t)
    =
    \sum_{l=0}^{d-1}[R_j(t)]_{lk}\hat a_l=\frac{1}{d}\sum_{r=0}^{d-1}\alpha_{r}^{(j)}(t)\hat{a}_{[k-r]_{d}}.
    \label{eq_U}
\end{equation}
Since $\hat H_{\mathrm{cyc}}|\mathrm{vac}\rangle=0$,
the vacuum is invariant under $\hat U_j(t)$.
Consequently, its action on an arbitrary Fock basis state is
\begin{equation}
    \hat U_j(t)| n_0,\ldots,n_{d-1}\rangle
    =
    \prod_{k=0}^{d-1}
    \frac{1}{\sqrt{n_k!}}\left[
            \sum_{l=0}^{d-1}
            [R_j(t)]_{lk}\hat a_l^\dagger
        \right]^{n_k}
    |\mathrm{vac}\rangle.
    \label{eq:Uj_Fock_action}
\end{equation}
Together with Eq.~\eqref{eq:Ueff_general}, this specifies the
full effective evolution at arbitrary times. At $t=\tau_d$, the relation
$R_j(\tau_d)=\mathcal M_d^{m_j}$ reduces
Eq.~\eqref{eq_U} to
\begin{equation}
    \hat U_j^\dagger(\tau_d)\hat a_k
    \hat U_j(\tau_d)
    =
    \hat a_{[k+m_j]_d}.
    \label{eq:conditional_operator_routing}
\end{equation}
We define the bosonic cyclic permutation operator $\hat P_d$ by $\hat P_d^\dagger\hat a_k\hat P_d =\hat a_{[k+1]_d}$ and $\hat P_d\hat a_k\hat P_d^\dagger =\hat a_{[k-1]_d}$. Because each $m_j=j-s$ is an integer,
$\hat U_j(\tau_d)$ and $\hat P_d^{m_j}$ have identical actions
on the mode operators. They therefore
coincide on the entire bosonic Fock space as $\hat U_j(\tau_d)=\hat P_d^{m_j}$. The resulting joint evolution is thus the qudit-controlled
cyclic permutation
\begin{equation}
    \hat U(\tau_d)
     =
    \sum_{j=0}^{d-1}
    | j\rangle\langle j|\otimes\hat P_d^{j-s}.
    \label{eq:Ueff_controlled_permutation}
\end{equation}
In particular, defining
$| \mathcal{N}_d^{(k)}\rangle=(\hat a_k^\dagger)^N/\sqrt{N!}
|\mathrm{vac}\rangle$, we obtain
\begin{equation}
    \hat U(\tau_d)
    \bigl(| j\rangle\otimes| \mathcal{N}_{d}^{(k)}\rangle\bigr)
    =
    | j\rangle\otimes
    | \mathcal{N}_{d}^{([k-m_j]_d)}\rangle.
    \label{eq:conditional_Fock_routing}
\end{equation}
Equation~(\ref{eq:conditional_Fock_routing}) makes the
controller-dependent chiral dynamics explicit. For $m_j=+1$, the
bosonic excitation undergoes a one-mode cyclic shift in one
direction at each routing time
\begin{equation}
\begin{aligned}
|\mathcal N_d^{(k)}\rangle
&\xrightarrow{\tau_d}
|\mathcal N_d^{([k-1]_d)}\rangle
\xrightarrow{\tau_d}
|\mathcal N_d^{([k-2]_d)}\rangle
\longrightarrow\cdots\\
&\xrightarrow{\tau_d}
|\mathcal N_d^{([k-(d-1)]_d)}\rangle
\xrightarrow{\tau_d}
|\mathcal N_d^{(k)}\rangle.
\end{aligned}
\label{eq:positive_chiral_cycle}
\end{equation}
For $m_j=0$, the evolution is frozen and the bosonic state remains
unchanged. In contrast, for $m_j=-1$, the excitation propagates in
the opposite direction
\begin{equation}
\begin{aligned}
|\mathcal N_d^{(k)}\rangle
&\xrightarrow{\tau_d}
|\mathcal N_d^{([k+1]_d)}\rangle
\xrightarrow{\tau_d}
|\mathcal N_d^{([k+2]_d)}\rangle
\longrightarrow\cdots\\
&\xrightarrow{\tau_d}
|\mathcal N_d^{([k+(d-1)]_d)}\rangle
\xrightarrow{\tau_d}
|\mathcal N_d^{(k)}\rangle.
\end{aligned}
\label{eq:negative_chiral_cycle}
\end{equation}
More generally, a controller state with eigenvalue $m_j$
produces a cyclic displacement by $m_j$ modes during each interval
$\tau_d$, so that at $t=n\tau_d$
\begin{equation}
|\mathcal N_d^{(k)}\rangle
\longrightarrow
|\mathcal N_d^{([k-nm_j]_d)}\rangle.
\label{eq:general_chiral_routing}
\end{equation}
Thus, positive and negative values of $m_j$ select opposite chiral
routing directions, while $m_j=0$ defines a stationary branch.
Since $m_j=-s,-s+1,\ldots,s$, the $d$ controller states generate
$d$ distinct cyclic displacements at the common routing time
$\tau_d$, mapping any fixed input mode onto all $d$ output modes
exactly once.

\section{General preparation of odd-$d$-mode NOON states}
\label{Sec:III}

\begin{figure}[t]
\centering
\includegraphics[width=0.95\columnwidth]{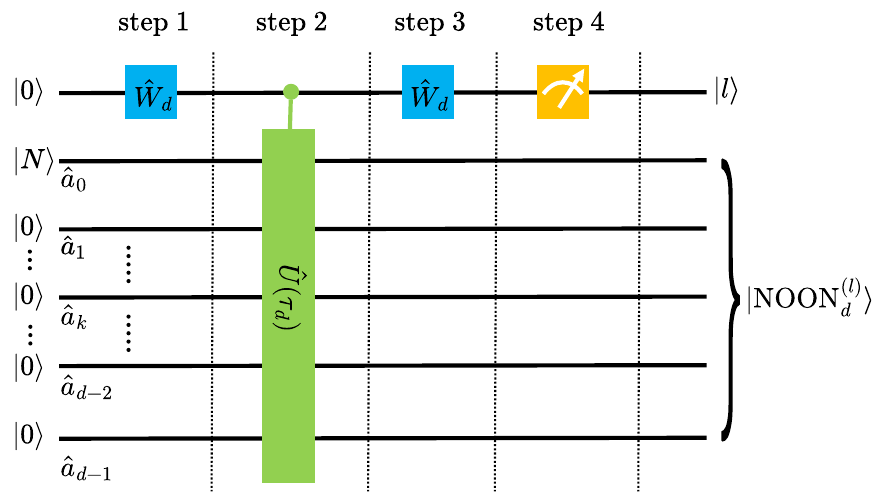}
\caption{Schematic diagram of the quantum circuit for generating the $d$-mode NOON states. It consists of the WH gate applied to the $d$-level system, the $d$-level system-controlled time evolution of the bosonic modes under $\hat{U}(\tau_d)$, a second application of the WH gate to the $d$-level system, and finally a projective measurement performed on the $d$-level system to generate the NOON state.}
\label{Fig.2}
\end{figure}

Having established the common‑time controlled routing in Sec.~\ref{Sec:IIB}, we now show that it directly provides a general protocol for preparing odd‑$d$-mode NOON states. We define
\begin{equation}
|\mathcal{N}_{d}^{(k)}\rangle
\equiv
|0,\ldots,0,
\underset{k{\rm th}}{N},
0,\ldots,0\rangle.
\end{equation}
We assume that the bosonic excitation is initially localized in the first mode and prepare the controller in a basis state $\ket{0}$. The initial state of the system is $|\Psi_{0}\rangle=|0\rangle\otimes|\mathcal{N}_{d}^{(0)}\rangle$. 

The entire process can be divided into four steps, denoted by step 1, step 2, step 3 and step 4 (as shown in Fig. \ref{Fig.2}). In the following, we will introduce our protocol step by step. In step 1, we prepare an equal superposition of the $d$ controller states. We introduce the $d$-dimensional Walsh-Hadamard (WH) gate \cite{PhysRevResearch.5.043087,PhysRevA.111.012623}
\begin{equation}
\hat W_d
=
\frac{1}{\sqrt d}
\sum_{j_{1},j_{2}=0}^{d-1}
q_d^{j_{1}j_{2}}
\ket{j_{1}}\bra{j_{2}},
\end{equation}
where $q_d=e^{i2\pi/d}$. The high-fidelity WH gate has been reported in Refs. \cite{PhysRevLett.125.180504,PhysRevLett.126.210504}. Acting $\hat{W}_{d}$ on the initial controller state, one obtain
\begin{equation}
\ket{\Psi^{(d)}_1}
=
\frac{1}{\sqrt d}
\sum_{j=0}^{d-1}
\ket{j}\otimes|\mathcal{N}_{d}^{(0)}\rangle.
\label{eq:psi1}
\end{equation}

In step 2, we apply $\hat{U}(\tau_{d})$ on state $\ket{\Psi_{1}}$, which yields
\begin{equation}
\ket{\Psi^{(d)}_2}
=
\frac{1}{\sqrt d}
\sum_{j=0}^{d-1}
\ket{j}\otimes
|\mathcal{N}_{d}^{([s-j]_{d})}\rangle.
\label{eq:psi2}
\end{equation}
We also used the relations $\hat{P}_d\hat{a}_k\hat{P}_d^\dagger=\hat{a}_{k-1}$ and $\hat{P}_d^{m_{j}}|\mathcal{N}_{d}^{(0)}\rangle=|\mathcal{N}_{d}^{([s-j]_d)}\rangle$. Equation~(\ref{eq:psi2}) represents a maximally correlated controller–bosonic state containing $d$ distinct routing branches.

In step 3, applying $\hat W_d$ once again to the controller gives
\begin{equation}
\ket{\Psi^{(d)}_3}
=
\frac{1}{\sqrt d}
\sum_{l=0}^{d-1}
q^{ls}\ket{l}\otimes
\ket{\mathrm{NOON}^{(l)}_{d}},
\label{eq:Psi3_general}
\end{equation}
where
\begin{equation}
\ket{\mathrm{NOON}^{(l)}_{d}}
=
\frac{1}{\sqrt d}
\sum_{j=0}^{d-1}
q_d^{-lj}|\mathcal{N}_{d}^{(j)}\rangle.
\label{eq:NOON}
\end{equation}
and the index $l$ is understood modulo $d$.
Since the $d$ states $\hat{P}_d^{ m_j}|\mathcal{N}_{d}^{(1)}\rangle$ are precisely the $d$ Fock states in which all $N$ excitations occupy one of the $d$ bosonic modes, Eq.~(\ref{eq:NOON}) is a $d$-mode NOON state. Moreover,
\begin{equation}
\langle\mathrm{NOON}^{(l')}_{d}|\mathrm{NOON}^{(l)}_{d}\rangle
=\frac{1}{d}\sum_{j=0}^{d-1}q_d^{(l'-l)j}
=\delta_{ll'},
\end{equation}
so the $d$ possible measurement outcomes correspond to a complete orthonormal Fourier NOON basis.

Finally, in step 4, a projective measurement of the controller
with outcome $l$ prepares
$\ket{\mathrm{NOON}^{(l)}_{d}}$. For the ideal protocol, each outcome occurs with probability $P_\ell=1/d$. All outcomes represent valid $d$-mode NOON states that differ only by known relative phases. They can therefore either be retained as distinct phase‑coded NOON states or converted to the same reference NOON state by outcome‑dependent phase feed‑forward.

The same conditional-routing protocol can also be extended to the
preparation of multimode entangled coherent states. We define the localized multimode coherent state
\begin{equation}
\ket{\mathcal C_d^{(k)}(\beta)}
\equiv
\ket{0,\ldots,0,
\underset{k{\rm th}}{\beta},
0,\ldots,0}.
\label{eq:localized_coherent_state}
\end{equation}
For an arbitrary odd dimension $d$, we may initialize the bosonic subsystem in $\ket{\mathcal C_d^{(0)}(\beta)}$, with the coherent state localized in a single mode. Following the same sequence of the WH operation, controlled cyclic routing, a second WH operation, and projective measurement, the measurement outcome $l$ prepares the phase-coded $d$-mode entangled coherent state
\begin{equation}
\ket{\mathrm{ECS}_{d}^{(l)}}
=
\frac{1}{\sqrt{M_l(\beta)}}
\sum_{j=0}^{d-1}
q_d^{-lj}
\ket{\mathcal C_d^{([s-j]_d)}(\beta)},
\label{eq:ECS_general}
\end{equation}
where $M_l(\beta)=d[1-e^{-|\beta|^2}+de^{-|\beta|^2}\delta_{l0}]$. In particular, the
equal-phase branch takes the form
\begin{equation}
\ket{\mathrm{ECS}_{d}^{(0)}}=\frac{1}{\sqrt{M_0(\beta)}}
\sum_{k=0}^{d-1}
\ket{\mathcal C_d^{(k)}(\beta)}.
\label{eq:ECS_equal_phase}
\end{equation}
For sufficiently large $|\beta|$, the different coherent-state
branches become nearly orthogonal \cite{PhysRevA.45.6811,PhysRevLett.107.083601,
PhysRevA.110.L010602,Israel:19}. The three-mode entangled coherent
state discussed previously is recovered by setting $d=3$.

The protocol is not restricted to equal probability amplitudes. Instead of the equal superposition in Eq.~(\ref{eq:psi1}), we introduce a general qudit state-preparation
unitary $\hat V$ satisfying
\begin{equation}
\hat V\ket{0}
=
\sum_{j=0}^{d-1}
c_j\ket{j},
\end{equation}
with $\sum_{j=0}^{d-1}|c_j|^2=1$. During the subsequent conditional-routing stage, each controller
component $\ket{j}$ routes the initial bosonic excitation from mode
$0$ to the mode $[s-j]_d$, while preserving its amplitude $c_j$.
The second WH operation then coherently mixes the controller states
and attaches the outcome-dependent phase factor $q_d^{lj}$ to the
$j$th routing branch. Consequently, upon projecting the controller
onto the computational state $\ket{l}$, the conditional bosonic state
is
\begin{equation}
|\psi_l\rangle
=
\sum_{j=0}^{d-1}
c_ {j}
q_d^{l j}
|\mathcal{N}_{d}^{[s-j]_d}\rangle,
\end{equation}
up to normalization. Because the routing branches are mutually orthogonal, the measurement probabilities $P_l$ are irrelevant to $c_{j}$, while the known phase factors $q_d^{l j}$ can be compensated by feed‑forward operations. Hence the same conditional‑routing architecture permits the preparation of odd‑$d$-mode NOON‑type states with arbitrary complex amplitudes.
Such weighted multimode superpositions are of particular interest
for quantum metrology, where unequal branch amplitudes can be
tailored to different estimation tasks
\cite{389g-d3sy}.

\section{Protocol for three- and five-mode NOON states}
\label{Sec:IV}

The general construction developed in Secs.~\ref{Sec:II} and \ref{Sec:III} applies to an arbitrary odd number $d=2s+1$ of bosonic modes. In this section, we illustrate the general results with the two lowest nontrivial cases, $d=3$ and $d=5$. The former realizes the controlled cyclic permutation with a single Floquet harmonic, whereas the latter is the first case for multiple Floquet harmonics.

\subsection{Three-mode NOON states protocol}

Related three-mode chiral-transfer schemes based on a two-level
auxiliary system have been investigated in Refs. \cite{PhysRevA.105.042422,PhysRevA.107.013702}. To prepare the three-mode NOON states, we consider the case of $d=3$ \cite{PhysRevA.106.L011301}, focusing on a hybrid quantum system composed of a three-level system (with energy levels denoted as $\ket{0}$, $\ket{1}$ and $\ket{2}$) coupled to three bosonic modes. For $d=3$, we have $s=1$. Therefore, a single Floquet harmonic is sufficient to realize the
exact three-mode cyclic permutation. From the general parameter
choice $G_p$, we immediately obtain $G_1=\sqrt{2\Omega\nu/3}$. The corresponding collective-spin transition amplitudes are $g_0^{(1)}=g_1^{(1)}=\sqrt{2}G_1$.

The effective hopping coefficient follows directly from
Eq.~\eqref{eq:Jr_closed}, $J_1=(\Omega/2)\csc\left(\pi/3\right)=\Omega/\sqrt{3}$. Accordingly, the effective Hamiltonian becomes
\begin{equation}
    \hat H_{\rm eff}^{(3)}
    =
    iJ_{1}\hat S_z
    \hat{\bm A}^{\dagger}
    \left(
    \mathcal M_3-\mathcal M_3^\dagger
    \right)
    \hat{\bm A},
    \label{eq:Heff_three}
\end{equation}
where $\hat{\bm A}=(\hat a_1,\hat a_2,\hat a_3)^\mathsf T$.
The common cyclic-transfer time is $\tau_3=2\pi/(3\Omega)$, at which $e^{-ih_{\rm cyc}^{(3)}\tau_3}=\mathcal M_3$. Consequently, the three controller eigenstates $m_j=-1,0,1$ generate $\mathcal M_3^{-1},I_3,\mathcal M_3$, respectively. Thus, the sign of $m_j$ determines the direction of the chiral transfer, while the $m_j=0$ component corresponds to dynamical freezing.

\begin{figure}[t]
\centering
\includegraphics[width=0.95\columnwidth]{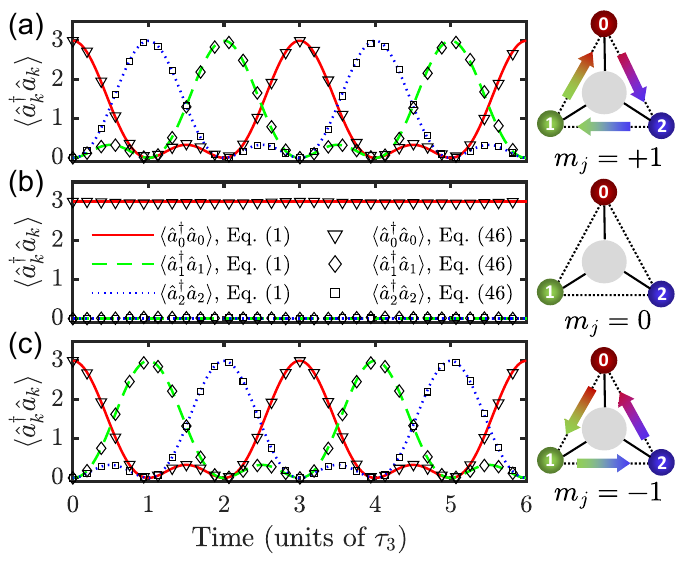}
\caption{Chiral dynamical properties of the excitation number distribution $\langle \hat{a}_{k}^{\dagger}\hat{a}_{k}\rangle$ for the three bosonic modes ($k=0,1,2$), studied via numerical simulation based on the Floquet-engineered Hamiltonian in Eq. (\ref{eq.ori_H}) and analytical evaluation using the effective Hamiltonian in Eq. (\ref{eq:Heff_three}). The three modes are initialized in the Fock state $\ket{3,0,0}$ while the three-level system is prepared in the (a) $\ket{2}$ state, (b) $\ket{1}$ state, and (c) $\ket{0}$ state. The red, green and blue lines correspond to bosonic modes 0, 1 and 2, respectively. The parameters are set as $\omega_{a}/2\pi=\omega_{q}/2\pi=8$ GHz, $G_{1}/2\pi=40$ MHz, $\nu/2\pi=1.2$ GHz, and $\varphi_{k}=2\pi k/3$.}
\label{Fig.3}
\end{figure}

Our analysis reveals that the propagation direction of the initially prepared state (clockwise, dynamical freezing, or counterclockwise) is determined by the state of the three-level system. This chiral transport originates from the broken time-reversal symmetry of the system. To intuitively characterize the nonreciprocal chiral excitation transport dynamics regulated by $m_{s}$, we adopt the parameter settings $\omega_{a}/2\pi=\omega_{q}/2\pi=8$ GHz, $G_{1}/2\pi=40$ MHz, $\nu/2\pi=1.2$ GHz, $\varphi_{k}=2\pi k/3$, and plot the temporal evolution of the excitation number population $\langle \hat{a}_{k}^{\dagger}\hat{a}_{k}\rangle$ for the initial states of $\ket{j}\otimes\ket{3,0,0}$ in Fig. \ref{Fig.3}, which is obtained from numerical simulation of the system Hamiltonian in Eq. (\ref{eq.ori_H}) and analytical solution of the effective Hamiltonian in Eq. (\ref{eq:Heff_three}). Here, Fig. \ref{Fig.3} (a), (b) and (c) correspond to clockwise chiral evolution, dynamical freezing and counterclockwise chiral evolution, respectively. The numerical results show excellent agreement with the predictions of the effective Hamiltonian in the parameter regime considered here. Consistent with the above analysis, the chiral direction is uniquely determined by the state of the three-level system. Notably, the propagation period of chiral states is independent of the three-level state and is solely determined by the parameters $G_{1}$ and $\nu$.


Here, we elaborate on four steps for generation of three-mode NOON states. Specifically, we assume the initial state is prepared in $\ket{\Psi(0)}=\ket{0}\otimes|\mathcal{N}_{3}^{(0)}\rangle$ where the three-level system is prepared in the $\ket{0}$ state and the bosonic system is in the Fock state $|\mathcal{N}_{3}^{(0)}\rangle$. The WH gate for a three-level system is defined as 
\begin{equation}
    \hat{W}_3=\frac{1}{\sqrt{3}}\begin{bmatrix}
        1 & 1 & 1\\
        1 & q_3 & q_3^{2}\\
         1 & q_3^{2} & q_3
    \end{bmatrix},
\end{equation}
where $q_3=\exp{(i2\pi/3)}$ denotes the primitive cube root of unity.

In step 1, we apply $\hat{W}_3$ to the three-level system to generate a superposition of $\ket{0}$, $\ket{1}$ and $\ket{2}$. The final state of step 1 is as follows
\begin{equation}\label{eq.step1}
    \ket{\Psi^{(3)}_1}= \frac{1}{\sqrt{3}}\left(q_{3}\ket{0}+q_{3}^{2}\ket{1}+\ket{2}\right) \otimes |\mathcal{N}_{3}^{(0)}\rangle.
\end{equation}

In step 2, we activate the conditional dynamics, which is given by the effective Hamiltonian in Eq. (\ref{eq:Heff_three}). By applying the effective Hamiltonian in Eq. (\ref{eq:Heff_three}) for a duration of $\tau_{3}$, the bosonic states are evolved to different states dependent on the three-level system state. At the end of step 2, we obtain the following highly entangled qutrit–bosonic state \cite{PhysRevLett.116.220502}
\begin{small}
\begin{equation}\label{eq.step2}
    \ket{\Psi^{(3)}_2}=\frac{1}{\sqrt{3}}
    \left(q_{3} \ket{0} \otimes |\mathcal{N}_{3}^{(1)}\rangle + q_{3}^{2}\ket{1} \otimes |\mathcal{N}_{3}^{(0)}\rangle+ \ket{2} \otimes |\mathcal{N}_{3}^{(2)}\rangle\right).
\end{equation}
\end{small}

In step 3, the WH gate $\hat{W}_3$ is applied again to the auxiliary three-level system. Then we obtain the following entangled state
\begin{small}
\begin{equation}
 \ket{\Psi^{(3)}_3} = \frac{1}{\sqrt{3}}
 (\ket{0} \otimes \ket{\mathrm{NOON}_{3}^{(0)}} + \ket{1} \otimes \ket{\mathrm{NOON}_{3}^{(1)}}+ \ket{2} \otimes \ket{\mathrm{NOON}_{3}^{(2)}}),
\end{equation}
\end{small}
where
\begin{equation}
\begin{aligned}
         &\ket{\mathrm{NOON}_{3}^{(0)}}=\frac{1}{\sqrt{3}}\left(q_{3}|\mathcal{N}_{3}^{(0)}\rangle+q_{3}^{2}|\mathcal{N}_{3}^{(1)}\rangle+|\mathcal{N}_{3}^{(2)}\rangle\right),\\
         &\ket{\mathrm{NOON}_{3}^{(1)}}=\frac{1}{\sqrt{3}}\left(|\mathcal{N}_{3}^{(0)}\rangle+|\mathcal{N}_{3}^{(1)}\rangle+|\mathcal{N}_{3}^{(2)}\rangle\right),\\
         &\ket{\mathrm{NOON}_{3}^{(2)}}=\frac{1}{\sqrt{3}}\left(q_{3}^{2}|\mathcal{N}_{3}^{(0)}\rangle+q_{3}|\mathcal{N}_{3}^{(1)}\rangle+|\mathcal{N}_{3}^{(2)}\rangle\right).
\end{aligned}
\end{equation}

In step 4, we perform a projective measurement on the three-level system in the basis $\{\ket{0},\ket{1},\ket{2}\}$ for the state $\ket{\Psi^{(3)}_3}$ \cite{2506.17797}, causing the bosonic system to collapse onto distinct ideal three-mode NOON states \cite{singh2025}. Through the measurement of the three-level system, we obtain the following outcomes with equal probability: $\ket{0}$, $\ket{1}$, and $\ket{2}$. Correspondingly, the bosonic system collapses onto the respective NOON states: $\ket{\mathrm{NOON}_{3}^{(0)}}$, $\ket{\mathrm{NOON}_{3}^{(1)}}$, and $\ket{\mathrm{NOON}_{3}^{(2)}}$. Based on the protocol elaborated above, we achieve the controllable preparation of three-mode NOON states.

\subsection{Five-mode NOON-state protocol}
\label{Sec:IVD}

We now consider the first nontrivial realization beyond the
three-mode single-harmonic architecture. According to the general
construction in Sec.~\ref{Sec:II}, the five-mode case corresponds to
$d=5$ (i.e., $s=2$). Therefore, two independent Floquet harmonics
are required to reach the minimum rank $4$ needed for an
exact five-site cyclic permutation. The auxiliary system is
described by the spin-$2$ manifold and its four adjacent transition amplitudes take the form
\begin{equation}
    \left(
    g_1^{(p)},g_2^{(p)},g_3^{(p)},g_4^{(p)}
    \right)
    =
    G_p
    \left(
    2,\sqrt{6},\sqrt{6},2
    \right),
    \quad p=1,2.
    \label{eq:g5}
\end{equation}
For the two harmonics, the amplitudes introduced in Sec.~\ref{Sec:IIA}
become $G_1=\sqrt{2\Omega\nu/5}$, $G_2=2\sqrt{2\Omega\nu/5}$.
To first order in the high-frequency expansion, the effective
five-mode Hamiltonian is
\begin{equation}
\hat{H}_{\rm eff}^{(5)}=i \hat{S}_z\hat{\bm A}^\dagger\left[J_1 \left(\mathcal{M}_5 - \mathcal{M}_5^{-1} \right)+ J_2 \left(\mathcal{M}_5^2 - \mathcal{M}_5^{-2} \right)\right]\hat{\bm A},
\label{eq:Heff5}
\end{equation}
with $\hat{\bm A} = (\hat{a}_1, \hat{a}_2, \hat{a}_3, \hat{a}_4, \hat{a}_5)^\mathsf T$. The two effective hopping amplitudes are $J_1=(\Omega/2)\csc(\pi/5)$, $J_2=-(\Omega/2)\csc(2\pi/5)$.
Thus, unlike the three-mode ring, the five-mode network requires
both nearest- and next-nearest-neighbor chiral hoppings.
At the common evolution time $\tau_5=2\pi/(5\Omega)$,
the bosonic evolution realizes one cyclic permutation $e^{-i\hat H_{\rm cyc}^{(5)}\tau_5}=\hat P_5$.
The controlled evolution takes the
standard form
\begin{equation}
    \hat U(\tau_5)
    =
    \sum_{j=0}^{4}
    |j\rangle\langle j|
    \otimes
    \hat P_5^ {m_j}.
    \label{eq:UCP5}
\end{equation}

\begin{figure}[t]
\centering
\includegraphics[width=0.9\columnwidth]{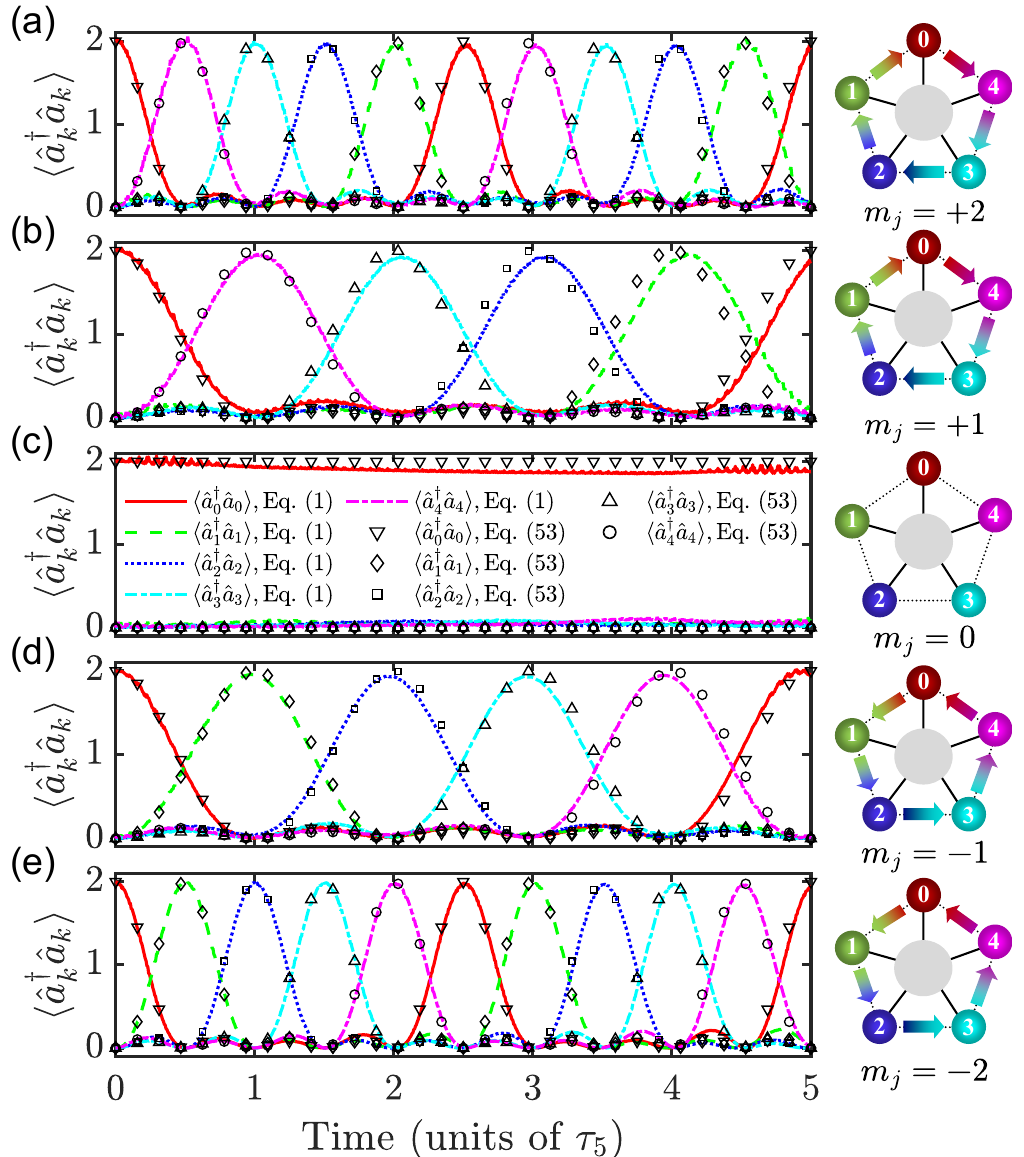}
\caption{Chiral dynamical properties of the excitation number distribution $\langle \hat{a}_{k}^{\dagger}\hat{a}_{k}\rangle$ for the five bosonic modes ($k=0,1,2,3,4$), studied via numerical simulation based on the Floquet-engineered Hamiltonian in Eq. (\ref{eq.ori_H}) and analytical evaluation using the effective Hamiltonian in Eq. (\ref{eq:Heff5}). The five modes are initialized in the Fock state $\ket{2,0,0,0,0}$ while the five-level system is prepared in the (a) $\ket{4}$ state, (b) $\ket{3}$ state, (c) $\ket{2}$ state, (d) $\ket{1}$ state, and (e) $\ket{0}$ state. The red, green, blue, cyan and magenta lines correspond to bosonic modes 0, 1, 2, 3 and 4, respectively. Here $\varphi_{k}=2\pi k/5$, and other parameters are the same as those in Fig. \ref{Fig.3}.}
\label{Fig.4}
\end{figure}

To further illustrate the conditional chiral dynamics in the
five-mode configuration, Fig.~\ref{Fig.4} shows the time evolution
of the excitation-number distributions
$\langle \hat a_k^\dagger \hat a_k\rangle$ ($k=0,\ldots,4$)
for different initial states of the five-level controller.
According to $m_j=j-2$, the controller states
$\ket{0}$, $\ket{1}$, $\ket{2}$, $\ket{3}$, and $\ket{4}$
correspond to $m_j=-2,-1,0,+1,+2$, respectively.
As predicted by Eq.~(\ref{eq:group_velocity}), the sign of $m_j$
determines the direction of the chiral excitation flow, whereas
$|m_j|$ controls its propagation rate. Consequently, the
$m_j=\pm2$ branches evolve at twice the rate of the $m_j=\pm1$ branches, while the $m_j=0$ branch remains stationary.

At the common routing time $\tau_5$, the different controller
sectors realize the corresponding cyclic permutations
$\hat P_5^{m_j}$. Thus, the $m_j=\pm1$ sectors produce a one-mode
cyclic displacement, whereas the $m_j=\pm2$ sectors yield a
two-mode net displacement. The numerical results obtained from the
original time-dependent Hamiltonian in Eq.~(\ref{eq.ori_H}) agree
well with the dynamics predicted by the effective Hamiltonian in
Eq.~(\ref{eq:Heff5}), confirming the validity of the
Floquet-engineered description for the five-mode configuration.

We next apply the general NOON-state protocol of Sec.~\ref{Sec:III}.
In
accordance with the three-mode realization, we initialize the
five-level controller in its lowest spin state, $\ket{\Psi^{(5)}(0)}=\ket{0}\otimes|\mathcal{N}_{5}^{(0)}\rangle$.
Applying the five-dimensional WH gate gives
\begin{equation}
\ket{\Psi^{(5)}_1}
=\left(\frac{1}{\sqrt5}
\sum_{j=0}^{4}
\ket{j}\right)\otimes|\mathcal{N}_{5}^{(0)}\rangle,
\label{eq:five_step1}
\end{equation}
where $q_5=e^{i2\pi/5}$. The phase factors in
Eq.~(\ref{eq:five_step1}) are fixed by the initial controller state
and do not affect the equal weights of the five routing branches.
After the controlled cyclic routing, the state becomes
\begin{equation}
\ket{\Psi^{(5)}_2}
=
\frac{1}{\sqrt 5}
\sum_{j=0}^{4}
\ket{j}\otimes
|\mathcal{N}_{d}^{([2-j]_{5})}\rangle.
\label{eq:Psi5route}
\end{equation}
The five controller components are therefore correlated with
five mutually orthogonal bosonic output modes.
Finally, applying $\hat W_5$ to the controller once again gives
\begin{equation}
\ket{\Psi^{(5)}_3}
=
\frac{1}{\sqrt5}
\sum_{l=0}^{4}
q_5^{2l}\ket{l}\otimes
\ket{\mathrm{NOON}_{5}^{(l)}},
\label{eq:five_step3}
\end{equation}
where the NOON-state index is understood modulo $5$. Performing a projective measurement in the computational basis
gives the five-mode NOON state
\begin{equation}
\begin{aligned}
\ket{\mathrm{NOON}^{(l)}_{5}}
=
\frac{1}{\sqrt 5}
\sum_{j=0}^{4}
q_5^{-lj}|\mathcal{N}_{5}^{(j)}\rangle,
\label{eq:NOON5}
\end{aligned}
\end{equation}
where $q_5=e^{i2\pi/5}$, $l=0,1,2,3,4$.
Each measurement outcome occurs with probability $1/5$, and
the five resulting states form an orthonormal Fourier NOON
basis.

Having illustrated the general multi-harmonic construction with the $d=5$ example, we now return to the experimentally more accessible $d=3$ implementation and analyze its sensitivity against realistic imperfections in Sec. \ref{Sec:V}.

\section{Discussion on Systematic Errors}
\label{Sec:V}
In realistic implementations, deviations from ideal parameters and
environmental noise inevitably affect the performance of the
state-preparation protocol. To assess the sensitivity of the central conditional-routing operation, we take the three-mode realization as a representative
example and analyze the influence of several relevant imperfections,
including frequency mismatch, coupling-strength deviations,
higher-order and counter-rotating terms, as well as dissipation and
decoherence.

Since the conditional cyclic-routing operation constitutes the
central dynamical step of the protocol, we quantify its accuracy by
comparing the ideal state obtained after step 2 with the state evolved
under the corresponding nonideal Hamiltonian or master equation.
Specifically, starting from the state
$\ket{\Psi^{(3)}_1}$ in Eq.~(\ref{eq.step1}), we evolve the system for
the routing time $\tau_3$ and define the conditional-routing fidelity
as $\mathcal{F} = |\langle \tilde{\Psi}^{(3)}_2 |\Psi^{(3)}_2\rangle|^2$, where $\ket{\Psi^{(3)}_2}$ is the ideal three-mode
controller--bosonic state in Eq.~(\ref{eq.step2}), while
$\ket{\tilde{\Psi}^{(3)}_2}$ denotes the state obtained in the
presence of the imperfection under consideration. The following
subsections use this three-mode benchmark to identify the parameter
regimes in which the Floquet-engineered conditional routing remains
robust.

\subsection{Frequency mismatch between the three-level system and bosonic modes}
\label{Sec:VA}
\begin{figure}[t]
\centering
\includegraphics[width=0.9\columnwidth]{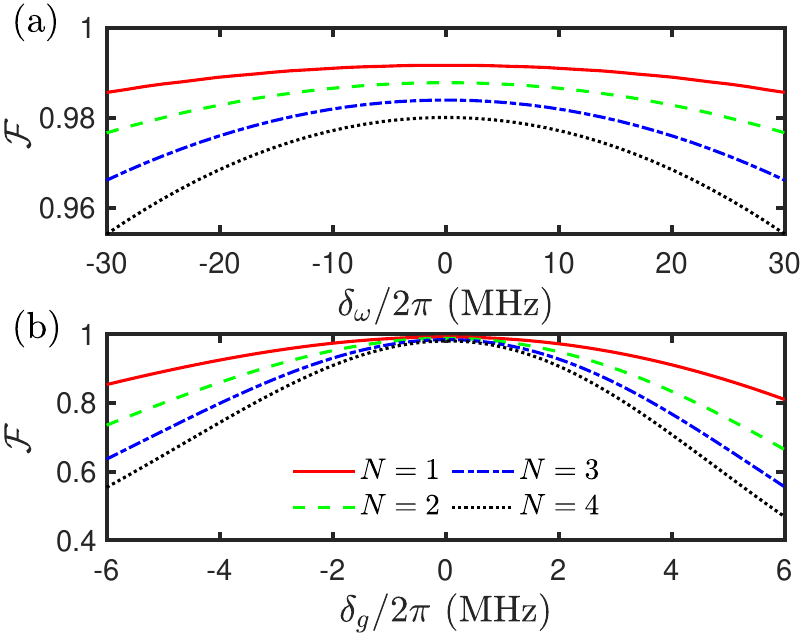}
\caption{Conditional-routing fidelity $\mathcal{F}$ as a function of (a) the frequency detuning $\delta_{\omega}/2\pi$ and (b) the coupling strength deviation $\delta_{g}/2\pi$ under the Hamiltonian with errors, for $N=1,2,3,4$. The parameter values here are the same as those in Fig. \ref{Fig.3}.}
\label{Fig.5}
\end{figure}

The resonant condition between the three bosonic modes and the three-level system was assumed in our previous discussions. In practice, however, it is difficult to achieve precise control over the system frequencies. We thus present a detailed analysis of the impact of frequency mismatch on the preparation of NOON states in this subsection.

Specifically, the frequencies of the bosonic modes and the three-level system satisfy $\omega_{a}-\omega_{q}=\delta_{\omega}$, where $\delta_{\omega}$ characterizes the strength of frequency detuning. Based on this detuning model, we perform numerical calculations of the fidelity of the target state as a function of the detuning parameter for $N=1,2,3,4$, with the results presented in Fig. \ref{Fig.5}(a). For a fixed $N$, it can be observed that the fidelity of the target state exhibits a monotonically decreasing trend with increasing detuning $|\delta_{\omega}|$. This arises because frequency detuning breaks the coherent resonant coupling between the three bosonic modes and the three-level system, suppresses efficient population transfer toward the target state, and consequently reduces the state fidelity. For a given detuning strength, a larger $N$ leads to a lower fidelity. This is because the coupling strength for the Fock state $\ket{N}$ scales up by a factor of $\sqrt{N}$, which causes the perturbative approximation to break down in the case of large $N$. 
Figure \ref{Fig.5}(a) clarifies the quantitative dependence of the conditional-routing fidelity on the frequency mismatch and therefore provides an estimate of the detuning tolerance of the routing operation. For $N=3$, the conditional-routing fidelity remains above 0.96 even at $|\delta_{\omega}|/2\pi=30~\rm{MHz}$, indicating that the routing dynamics is relatively insensitive to moderate frequency mismatch within the parameter regime considered here.

\subsection{Deviation of coupling strengths between bosonic modes and the three-level system}

The coupling strength is susceptible to variations in the size and position of the bosonic modes, as well as environmental perturbations. These effects can be incorporated by renormalizing the coupling strengths as $g_{1}^{(1)}=g_{2}^{(1)}=\sqrt{2}(G_{1}+\delta_{g})$ where $\delta_{g}$ quantifies the coupling strength error induced by the aforementioned effects.

To examine the effect of error in the coupling strength on the system fidelity, we plot the curves of the target state fidelity as a function of the deviation of coupling strength between the bosonic mode and the three-level system for $N=1,2,3,4$ in Fig. \ref{Fig.5}(b). It is seen that the fidelity decreases with increasing $|\delta_{g}|$, and the degradation rate differs for positive and negative coupling deviations. This behavior arises because a shift in the coupling strength $G_{1}$ modifies the effective coupling strength $J_1$, which in turn alters the evolution time $\tau_{3}$. Since we terminate the evolution at a fixed evolution time, the fidelity is inevitably reduced. In this case, a fidelity close to 1 can be readily recovered by simply adjusting the evolution time according to the actual coupling strength. Moreover, it can be observed that the fidelity of the target state is less robust against errors in the coupling strength than against frequency errors in Sec. \ref{Sec:VA}.
Quantitatively, for $N=3$, the conditional-routing fidelity remains above 0.9 for a relative coupling-strength deviation of approximately $6\%$. This result indicates that the conditional-routing operation retains a relatively high accuracy against moderate coupling fluctuations within the parameter range considered here.

\subsection{Effect of high-frequency terms}
In the Hamiltonian of our proposed scheme given in Eq. (\ref{eq.ori_H}), the counter-rotating terms have been neglected under the rotating-wave approximation (RWA), which requires the coupling strength to satisfy $g \ll \omega_{a}$. Furthermore, in deriving the effective Hamiltonian without counter-rotating terms in Eq. (\ref{eq.5}), we have also discarded several contributions from high-order correction terms. To further evaluate the influence of parameter values and the systematic errors originating from the effective Hamiltonian, we now take the counter-rotating terms into account to quantify the systematic errors introduced by the above two approximations concerning high-frequency terms and their impact on the fidelity of target state preparation. The system Hamiltonian can be written as
\begin{equation}\label{eq.25}
   \begin{aligned}
    \hat{H}'(t) =& \sum_{j=0}^{d-1} \omega_{q_{j}}\ket{j}\bra{j}+\sum_{k=0}^{d-1}\omega_{a}\hat{a}_{k}^{\dagger}\hat{a}_{k}\\
    &+\sum\limits_{j=0}^{d-2}\sum_{k=0}^{d-1}g_{jk}(t)(\ket{j}\bra{j+1}\hat{a}_{k}^{\dagger}+\ket{j}\bra{j+1}\hat{a}_{k}+\rm{H.c.}),   
    \end{aligned}
\end{equation}
where $\ket{j}\bra{j+1}\hat{a}_{k}$ and $\ket{j+1}\bra{j}\hat{a}_{k}^{\dagger}$ represent the counter-rotating terms. These terms introduce virtual transitions that cause the population to leak from the ideal three-state cyclic transfer to other Fock states. Their influence on the conditional-routing fidelity of the system is jointly regulated by the bosonic mode frequency $\omega_{a}$ and the driving frequency $\nu$.

To quantify the validity of the leading-order Floquet Hamiltonian,
we compare the coupling matrix elements of the oscillating terms with
their corresponding harmonic frequencies. In the truncated bosonic
Hilbert space containing at most $N_{\max}$ excitations, the largest
bosonic matrix element scales as $\sqrt{N_{\max}+1}$, while the
maximum collective-spin matrix element is of order $d/2$. A
sufficient condition for the high-frequency expansion is therefore
\begin{equation}
\epsilon_{\mathrm{HF}}
\equiv
\max_{p,j}
\frac{
G_p\sqrt{j(d-j)}\sqrt{N_{\max}+1}
}{
p\nu
}
\ll1.
\label{eq:HF_condition_general}
\end{equation}
Using
$\max_j\sqrt{j(d-j)}\simeq d/2$, we obtain
\begin{equation}
\epsilon_{\mathrm{HF}}
\lesssim
\sqrt{
\frac{d(N_{\max}+1)\Omega}{2\nu}
}
\ll1.
\label{eq:HF_condition_simplified}
\end{equation}
Equation~(\ref{eq:HF_condition_simplified}) shows that increasing
either the number of modes or the accessible excitation number
places a more stringent requirement on the driving frequency.
Likewise, a larger target routing frequency $\Omega$ shortens the
routing time but increases the relative importance of higher-order
Floquet corrections.

\begin{figure}[t]
\centering
\includegraphics[width=0.9\columnwidth]{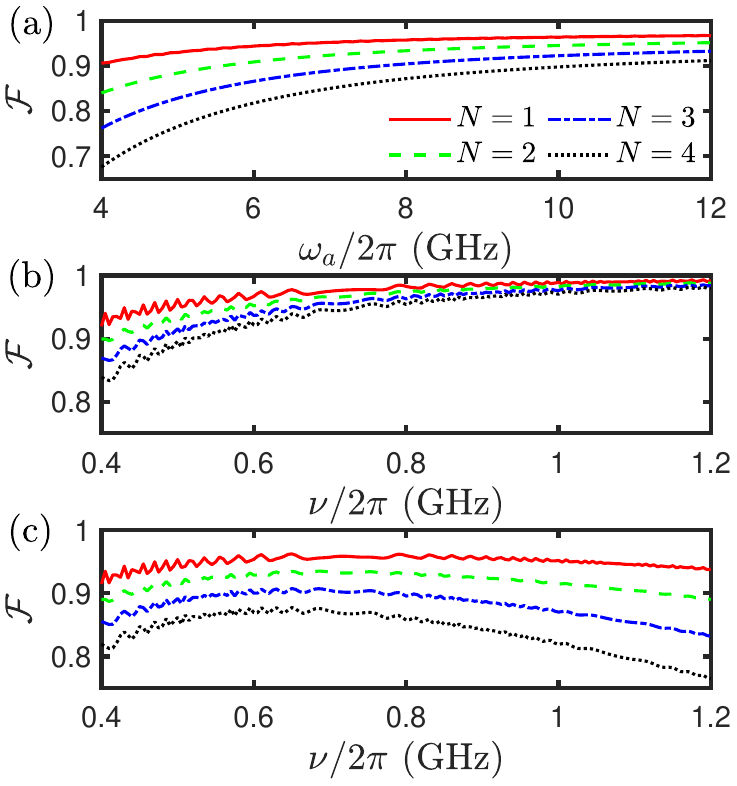}
\caption{Conditional-routing fidelity $\mathcal{F}$ under different Hamiltonians and system parameters for $N=1,2,3,4$. (a) Fidelity $\mathcal{F}$ from the evolution governed by the full Hamiltonian $\hat{H}'$ in Eq. (\ref{eq.25}) as a function of the bosonic mode frequency $\omega_{a}/2\pi$ for driving frequency $\nu/2\pi=0.7~\rm{GHz}$. (b) Fidelity $\mathcal{F}$ from the evolution governed by the Hamiltonian $\hat{H}(t)$ in Eq. (\ref{eq.ori_H}) as a function of the driving frequency $\nu/2\pi$ for bosonic mode frequency $\omega_{a}/2\pi=8~\rm{GHz}$. (c) Fidelity $\mathcal{F}$ from the evolution governed by the full Hamiltonian $\hat{H}'(t)$ in Eq. (\ref{eq.25}) as a function of the driving frequency $\nu/2\pi$ for bosonic mode frequency $\omega_{a}/2\pi=8~\rm{GHz}$. The rest of the parameter settings are the same as those in Fig. \ref{Fig.3}.}
\label{Fig.6}
\end{figure} 

To intuitively characterize the influence of high-frequency terms on the conditional-routing fidelity, we perform numerical simulations of the conditional-routing fidelity $\mathcal{F}$ based on different Hamiltonians and system parameters for the case of $N=1,2,3,4$, with the results shown in Fig. \ref{Fig.6}. Fig. \ref{Fig.6}(a) plots the fidelity $\mathcal{F}$ obtained from the evolution governed by the full Hamiltonian $\hat{H}'(t)$ in Eq. (\ref{eq.25}) as a function of the bosonic mode frequency $\omega_{a}/2\pi$ with the driving frequency fixed at $\nu/2\pi=0.7~\rm{GHz}$. As $\omega_{a}$ increases, the virtual transitions induced by the counter-rotating terms are gradually suppressed, leading to a monotonically increasing trend in the fidelity. At $\omega_{a}/2\pi=8~\rm{GHz}$, the fidelity exceeds 0.90 for $N=3$, indicating that the rotating-wave approximation becomes effectively valid under this parameter regime. When $\omega_{a}$ is further increased to $12~\rm{GHz}$, the fidelity rises above 0.93, verifying the beneficial effect of enlarging $\omega_{a}$ in suppressing errors arising from the counter-rotating terms. In contrast, a small $\omega_{a}$ makes the counter-rotating terms the dominant source of error, resulting in a drastic drop in fidelity and a severe deviation of the system dynamics from the ideal behavior.

To separate the influence of the counter-rotating terms from the inherent Floquet approximation, we also perform simulations in which these terms are omitted. Fig. \ref{Fig.6}(b) shows the fidelity obtained from the evolution governed by the Hamiltonian $\hat{H}(t)$ in Eq. (\ref{eq.ori_H}) as a function of driving frequency $\nu/2\pi$ with the bosonic mode frequency fixed at $\omega_{a}/2\pi=8~\rm{GHz}$. It is clearly shown that the conditional-routing increases monotonically with the driving frequency $\nu$ and approaches 1 in the high-frequency limit. This behavior is consistent with the Floquet expansion: as $\nu$ increases, the higher-order corrections become negligible, and the system dynamics converge to the ideal effective Hamiltonian.

Fig. \ref{Fig.6}(c) shows the fidelity obtained from the evolution governed by the full Hamiltonian $\hat{H}'$ in Eq. (\ref{eq.25}) as a function of the driving frequency $\nu/2\pi$ with the bosonic mode frequency fixed at $\omega_{a}/2\pi=8~\rm{GHz}$. The fidelity exhibits a nonmonotonic trend of first increasing and then decreasing, with an optimal driving frequency around $\nu/2\pi\approx0.7~\rm{GHz}$. For an excessively small $\nu$, the Floquet perturbation condition $\nu \gg g_{j}$ is no longer satisfied, and the superposition of high-order corrections and counter-rotating terms gives rise to low fidelity. As $\nu$ is gradually increased, the system enters the valid regime of Floquet perturbation, where errors from high-frequency terms are partially suppressed and the fidelity is improved. However, a further increase in $\nu$ suppresses the high-frequency terms further but reduces the effective coupling strength $J_1$, which in turn prolongs the evolution time $\tau_{3}$. This enhances the cumulative effect of high-frequency terms, while the weakened effective coupling reduces the strength of the system evolution, both of which lead to a decline in fidelity. These results show that the bosonic-mode frequency $\omega_{a}$ and the modulation frequency $\nu$ should be chosen by balancing the validity of the rotating-wave and high-frequency approximations against the resulting routing time. Thereby we can improve the accuracy of the conditional-routing dynamics within the parameter regime considered here. 

\subsection{Effect of dissipation}

To assess the influence of dissipation on the central conditional-routing stage, we numerically evaluate the conditional-routing fidelity within the framework of an open quantum system. Specifically, we incorporate the decoherence effects of each component in the hybrid system into the dynamical simulations, where the system dynamics is governed by the Lindblad master equation for the
density-matrix operator $\hat{\rho}(t)$ \cite{Scully_Zubairy_1997,Lindblad1976},
\begin{equation}\label{eq.26}
\begin{aligned}
    \frac{d\hat{\rho}(t)}{dt}=&-i[\hat{H}(t),\hat{\rho}(t)]+\sum_{j=0}^{1}\frac{\gamma}{2}\mathcal{L}[\ket{j}\bra{j+1}]\hat{\rho}(t)\\
    &+\sum_{k=0}^{2}\frac{\kappa}{2}\mathcal{L}[\hat{a}_{k}]\hat{\rho}(t).
\end{aligned}
\end{equation}
Here $\hat{H}(t)$ is given in Eq. (\ref{eq.ori_H}) and the Lindblad superoperator is defined as
\begin{equation}
    \mathcal{L}[\hat{O}]\hat{\rho}(t)=2\hat{O}\hat{\rho}(t) \hat{O}^{\dagger} - \hat{O}^{\dagger}\hat{O}\hat{\rho}(t) - \hat{\rho}(t) \hat{O}^{\dagger}\hat{O},
\end{equation}
in which $\hat{O}=\hat{a}_{k}$, $\ket{j}\bra{j+1}$ represent the decay operators with corresponding rates $\kappa=1/T_{1}$, and $\gamma=1/T_{1}$. Here $T_{1}$ is the same relaxation time of magnons and the three-level system. Since decoherence is present, the evolved state becomes a mixed state described by the density matrix $\hat{\rho}(t)$, while the target remains an ideal pure state $\ket{\Psi^{(3)}_2}$ in Eq. (\ref{eq.step2}). Then the fidelity can be calculated by $\mathcal{F}=\langle\Psi^{(3)}_2|\hat{\rho}(t)|\Psi^{(3)}_2\rangle$.

\begin{figure}[t]
\centering
\includegraphics[width=0.9\columnwidth]{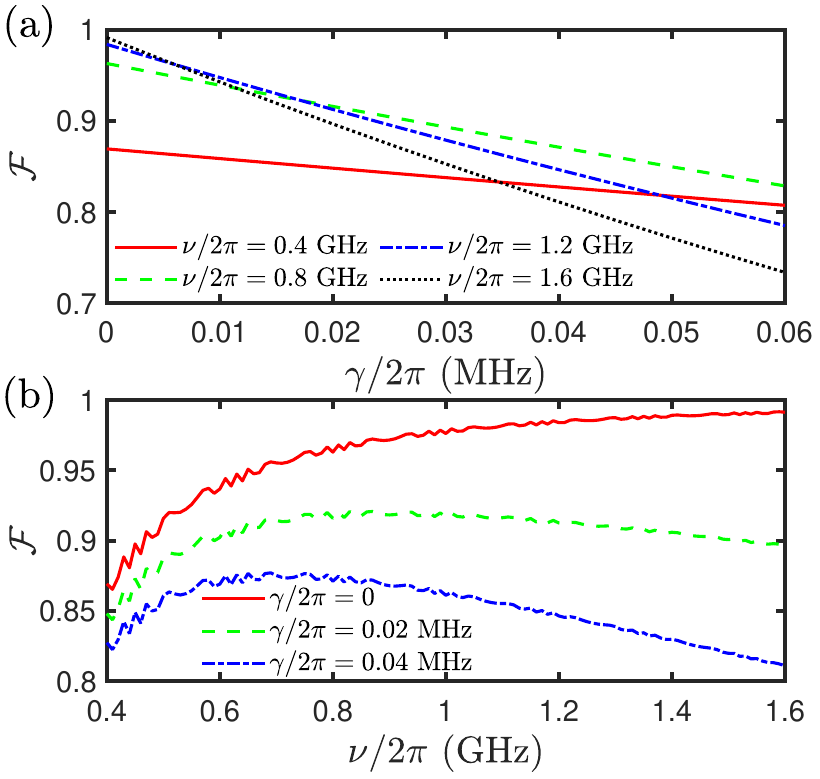}
\caption{Conditional-routing fidelity $\mathcal{F}$ under the master equation Eq. (\ref{eq.26}) as a function of (a) decoherence rate $\gamma/2\pi$ for different driving frequencies $\nu/2\pi$ and (b) driving frequency $\nu/2\pi$ for different decoherence rates $\gamma/2\pi$ at $N=3$. All other parameters are the same as those in Fig. \ref{Fig.3}.}
\label{Fig.7}
\end{figure}

In Fig. \ref{Fig.7}(a), we plot the fidelity $\mathcal{F}$ of the target state as a function of the decoherence rate $\gamma/2\pi$ for different driving frequencies $\nu/2\pi$ by using the master equation Eq. (\ref{eq.26}).  It can be seen that at a fixed $\nu$, the fidelity of the target state decreases with increasing $\gamma$. To analyze the influence of $\nu$ on the fidelity more clearly, we further plot the fidelity as a function of the driving frequency $\nu/2\pi$ for different decoherence rates $\gamma/2\pi$ in Fig. \ref{Fig.7}(b). Notably, the fidelity of the target state does not show a monotonically increasing trend with increasing $\nu$ when $\gamma\neq0$. This is because a larger $\nu$ can better satisfy the high-frequency approximation, whereas an excessively large $\nu$ reduces the effective coupling strength, thus prolonging the evolution time and exacerbating the effect of decoherence on the system. Therefore, it is necessary to strike a balance between the validity of the perturbative approximation and the effect of decoherence. 
Within the parameter range considered here, the conditional-routing dynamics retains a finite tolerance to dissipation. For example, at $\gamma/2\pi=0.04~\rm{MHz}$, the conditional-routing fidelity is approximately $0.85$ for the chosen parameters.

We emphasize that the fidelities discussed above characterize the
conditional-routing stage and should not be identified directly with
the fidelity of the final postselected NOON state. In a complete
description, the two controller WH operations and the final projective
measurement should also be included. For a measurement outcome $l$,
the final bosonic-state fidelity can be defined as
\begin{equation}
\mathcal{F}_l
=
\bra{\mathrm{NOON}_{d}^{(l)}}
\rho^{(l)}
\ket{\mathrm{NOON}_{d}^{(l)}},
\end{equation}
where $\rho^{(l)}$ denotes the normalized bosonic state conditioned
on outcome $l$. The present analysis assumes ideal controller gates
and projective measurements and therefore focuses on imperfections
accumulated during the conditional-routing dynamics.

\section{Experimental Feasibility}
\label{Sec:VI}

Based on the theoretical model of the coupling between a $d$-level system and $d$ magnon modes as well as the numerical calculation results presented earlier, we discuss the experimental feasibility of qutrit–bosonic state preparation by combining existing quantum experimental technologies. In terms of the experimental architecture, we propose a hybrid system combining superconducting circuits and magnon modes, which has emerged as a highly promising platform for quantum information processing \cite{RevModPhys.93.025005,PhysRevLett.113.083603}.

For an $d$-level system, we consider ($d-1$) identical two-level systems which serve as mediating couplers of bosonic modes \cite{PhysRevA.105.043704}. The required $d$-level subsystem can be implemented using the fully
symmetric Dicke manifold of $M=d-1$ identical two-level systems. We
identify the $j$th level as the symmetric Dicke state containing
$j-1$ excitations,
\begin{equation}
    |j\rangle
    \equiv
    |D_{j}^{(M)}\rangle
    =
    \frac{1}{\sqrt{\binom{M}{j}}}
    \sum_{\mathrm{perm}}
    |\underbrace{e\cdots e}_{j}
    \underbrace{g\cdots g}_{M-j}\rangle.
\end{equation}
The collective raising operator
$\hat \Sigma_{+}=\sum_{\mu=1}^{M}\hat{\sigma}_{\mu}^{+}$ then acts as
$\hat \Sigma_{+}|j\rangle=\sqrt{(j+1)(d-j-1)}|j+1\rangle$. Hence, the coupling
$g_j^{(p)}=G_p\sqrt{j(d-j)}$ arises naturally from collective Dicke
enhancement. The two-atom spin-$1$ system for $d=3$ is the simplest
realization of this construction. Multilevel spin degrees of freedom and macroscopic entanglement have been extensively investigated in spinor Bose--Einstein condensates, providing a natural platform for implementing higher-dimensional controller manifolds \cite{PhysRevA.73.023602}.

In particular, a three-level system consisting of two superconducting transmon qubits serves as the control unit, while the magnon modes hosted in three independent yttrium iron garnet (YIG) spheres act as the controlled bosonic modes. Specifically, the three-level system is constructed from the spin triplet states of two identical transmon qubits, corresponding to the energy levels $\ket{0}=\ket{00}, \ket{1}=(\ket{01}+\ket{10})/\sqrt{2}, \ket{2}=\ket{11}$, with energies $0$, $\omega_q$, and $2\omega_q$, respectively. The antisymmetric singlet $(\ket{01}-\ket{10})/\sqrt2$ is dark under a permutation-symmetric coupling and therefore does not participate in the dynamics. The frequency of each transmon is designed to be around $\omega_{q}/2\pi\approx8~\rm{GHz}$, which can be tuned by an external flux bias to ensure that the transition frequencies of the two qubits are equal, accurately matching the resonance frequency $\omega_{a}=\omega_{q}$ of the three magnon modes. This frequency range is within the optimal operating interval of both superconducting qubits and YIG magnon systems, which can be stably controlled by current technologies. The three magnon modes are realized by YIG spheres with diameters of approximately 0.3 mm, whose characteristic frequencies are tuned to resonate with the three-level system via an external magnetic field, and the inter-mode coupling is suppressed to a negligible level through structural optimization. The coupling between the magnons and the three-level system is achieved by microwave driving: each YIG sphere is capacitively coupled to both transmons, with the single-channel coupling strength set to $G_{1}$. For YIG spheres of the size employed in this work, the experimentally realized coupling strength $G_{1}/2\pi$ is approximately $10~\rm{MHz}$ \cite{Lachance-Quirion_2019,science.aaz9236,science.aaa3693}. By further optimizing the cavity filling factor, the coupling strength required by our protocol can be achieved in the near future.

In addition, the WH gate for the superconducting three-level qubit has been experimentally demonstrated with a duration of $35~\rm{ns}$ and a fidelity of $99.2\%$ \cite{PhysRevLett.125.180504}, providing a reliable tool for our quantum control protocol. 
The reported gate duration is much shorter than the typical coherence time of superconducting qubits $T_{1}$, which is about $40–60~\mu\rm{s}$ \cite{Yan2016,10.1063/1.5136262}, corresponding to the relaxation and dephasing rates about $16.67–25~\rm{kHz}$, suggesting that decoherence during the controller-gate operation need not be the dominant limitation.

Meanwhile, sufficiently low magnon dissipation is important for maintaining the accuracy of the conditional-routing dynamics. Previous experimental studies have shown that the selected YIG spheres possess an extremely low Gilbert damping coefficient $\alpha=2.7(5)\times10^{-5}$ \cite{10.1063/1.4977423}. For $\omega_{a}/2\pi=8~\rm{GHz}$, this damping corresponds to an intrinsic energy decay rate $\gamma=\alpha\omega_{a}\approx1.38~\rm{MHz}$, with a linewidth of $\gamma/2\pi\approx0.22~\rm{MHz}$ and a magnon lifetime of approximately $737~\rm{ns}$. This value is about one order of magnitude larger than $\gamma/2\pi=0.02~\rm{MHz}$ in our optimal simulations. However, the gap is not insurmountable. This factor can be realistically bridged by combining several advanced techniques: using ultra-high-purity raw materials and laser floating zone growth to reduce intrinsic magnon-phonon scattering \cite{PhysRevB.95.214423,cryst11010038}, annealing in oxygen to suppress impurities \cite{10.1063/1.4977423}, applying strain-engineered epitaxial shells to achieve $\alpha < 1\times10^{-5}$ \cite{PhysRevB.101.174431}, and employing chemical mechanical polishing plus atomic layer deposition coatings to eliminate surface defects and two-magnon scattering \cite{WANG20251007,LUO2024100841,OLIVEIRA2020166851,10.1063/5.0202639,PhysRevMaterials.6.044411}. Moreover, superconducting magnetic levitation can remove clamping-induced losses \cite{PhysRevA.108.063511}. These developments indicate possible routes toward reducing the magnon damping toward the regime considered in our numerical analysis. Reaching such a reduced-loss regime would improve the performance of the conditional-routing operation and narrow the gap between the idealized model and a practical implementation.

We also propose a hybrid architecture consisting of a superconducting giant atom coupled to three three-dimensional (3D) superconducting cavities as another feasible platform. In this architecture, the three-level system is realized by a transmon qutrit, and the three bosonic modes correspond to the same eigenmode of the 3D superconducting cavities. The 3D superconducting cavities provide an exceptionally low-dissipation environment, with quality factors exceeding $4\times10^{11}$ and photon lifetimes exceeding 2 s at a temperature of 10 mK, corresponding to an intrinsic decay rate of only about 0.5 Hz \cite{PhysRevApplied.13.034032}. Moreover, the giant-atom configuration enables engineerable dissipation via quantum interference, with effective decay rates tunable over a range of several MHz \cite{xiao2025}. This combination of a low intrinsic loss rate and tunable effective coupling makes the giant-atom–3D-cavity architecture a promising candidate for implementing the conditional-routing dynamics considered here. In particular, the reported cavity-loss parameters are compatible with the low-dissipation regime assumed in our numerical analysis.

\section{Conclusion}
\label{Sec:VII}

In summary, we have developed a Floquet-engineered conditional-routing framework for the preparation of multimode NOON states in bosonic networks. By coupling $d$ bosonic modes to a $d$-level controller, we obtain a state-dependent effective chiral Hamiltonian that enables the controller to coherently select different cyclic-routing branches. The restriction to odd $d$ follows from the structure of the effective coupling matrix. Each modulation harmonic contributes a matrix of rank at most two, whereas a generator of an exact $d$-mode cyclic permutation must have rank at least $d-1$. Moreover, the purely imaginary, antisymmetric circulant generator constructed here has a zero eigenvalue for the uniform Fourier mode. For even $d$, the alternating Fourier mode is also necessarily a zero mode, limiting the rank to at most $d-2$. Its evolution therefore cannot reproduce the eigenvalue $-1$ of the cyclic permutation on this mode. Consequently, adding further harmonics within the same coupling structure cannot realize an exact even-$d$ cyclic permutation; an additional interaction that lifts this zero mode is required. Based on this, we establish
a general protocol for preparing odd-$d$-mode NOON states, with
programmable relative amplitudes and phases through control of the
auxiliary system.

The three- and five-mode cases illustrate the two characteristic
regimes of the general framework. The three-mode realization requires
only a single harmonic and exhibits two oppositely directed chiral
branches together with a stationary branch, whereas the five-mode
case constitutes the first nontrivial multi-harmonic realization and
requires both nearest- and next-nearest-neighbor effective chiral
hoppings. We have also examined the sensitivity of the three-mode
protocol against representative imperfections and dissipation, and
discussed possible implementations in hybrid superconducting--bosonic
platforms. The reported experimental parameters of superconducting-qubit–magnon and giant-atom–3D-cavity systems indicate that the basic frequency, coupling, gate-control, and dissipation requirements of the conditional-routing architecture are compatible with experimentally relevant regimes, although a complete end-to-end implementation would additionally require accurate initialization, controller operations, and projective readout. The scheme also features is applicable to arbitrary odd $d$ within the coupling architecture considered here, and is compatible with various quantum experimental platforms \cite{wfj8-tgjz}. These results provide a scalable route to controllable
multimode entanglement generation based on Floquet-engineered chiral
dynamics, with potential applications in quantum information
processing and multiparameter quantum metrology.

An interesting direction for future work is to extend the same
conditional-routing architecture to more general definite
photon-number (DPN) states. By replacing the initially localized
Fock state with a more general fixed-total-number configuration
$\ket{n_1,\ldots,n_d}$, the controlled cyclic operation can generate
coherent superpositions of its cyclic permutations. Certain optimized DPN states have been shown to retain a metrological advantage in the presence of photon loss
\cite{PhysRevLett.102.040403,l44c-d76x}. Optimizing the present Floquet
architecture for loss-tolerant DPN-state generation will be explored
in future work.

\section*{Acknowledgments}
G.W. was supported by the Quantum Science and Technology-National Science and Technology Major Project (No. 2023ZD0300700). H.L. was supported by the National Natural Science Foundation of China (Grants No. 12575011) and the Science and Technology Development Plan Project of Jilin Province, China (Grant No. 20240101321JC). H.S. was supported by Science and Technology Development Plan Project of Jilin Province (Grant No. 20250102007JC), and National Natural Science Foundation of China under Grant No. 12274064.

\bibliography{ref}
\providecommand{\noopsort}[1]{}\providecommand{\singleletter}[1]{#1}%

\end{document}